\documentclass[journal=nalefd,manuscript=letter,layout=twocolumn,]{achemso}
\usepackage[sc]{mathpazo}
\usepackage[scaled=0.9]{helvet}
\usepackage[utf8]{inputenc}
\usepackage{graphicx}
\usepackage{amsmath,amssymb,amsfonts}
\usepackage{upgreek} 
\usepackage{textcomp}
\usepackage[export]{adjustbox} 
\usepackage[pdftex,dvipsnames]{xcolor}
\usepackage[pdftex]{hyperref}
\hypersetup{
	pdftitle={UV Photosensing Characteristics of Nanowire Based GaN/AlN Superlattices},
	colorlinks,
	citecolor=blue
	}
\usepackage{soul}
\usepackage{xspace}

\newcommand{\celsius}{$^{\circ}$C\xspace}

\title{UV Photosensing Characteristics of Nanowire Based GaN/AlN Superlattices}


\author{Jonas L\"ahnemann}
\affiliation{University Grenoble-Alpes, 38000 Grenoble, France}
\alsoaffiliation{CEA-Grenoble, INAC-PHELIQS, 17 av. des Martyrs, 38000 Grenoble, France}
\email{jonas.laehnemann@cea.fr}
\author{Martien Den Hertog}
\affiliation{University Grenoble-Alpes, 38000 Grenoble, France}
\alsoaffiliation{Institut Néel, CNRS, 25 av. des Martyrs, 38000 Grenoble, France}
\author{Pascal Hille}
\affiliation{I. Physikalisches Institut, Justus-Liebig-Universität Gießen, 35392 Gießen, Germany}
\author{Maria de la Mata}
\affiliation{Institut Català de Nanociència i Nanotecnologia (ICN2), CSIC and Barcelona Institute of Science and Technology, Campus UAB, Bellaterra, 08193 Barcelona, Catalonia, Spain}
\author{Thierry Fournier}
\affiliation{University Grenoble-Alpes, 38000 Grenoble, France}
\alsoaffiliation{Institut Néel, CNRS, 25 av. des Martyrs, 38000 Grenoble, France}
\author{J\"org Sch\"ormann}
\affiliation{I. Physikalisches Institut, Justus-Liebig-Universität Gießen, 35392 Gießen, Germany}
\author{Jordi Arbiol}
\affiliation{Institut Català de Nanociència i Nanotecnologia (ICN2), CSIC and The Barcelona Institute of Science and Technology (BIST), Campus UAB, Bellaterra, 08193 Barcelona, Catalonia, Spain}
\alsoaffiliation{Institució Catalana de Recerca i Estudis Avançats (ICREA), 08193 Barcelona, Catalonia, Spain}
\author{Martin Eickhoff}
\affiliation{I. Physikalisches Institut, Justus-Liebig-Universität Gießen, 35392 Gießen, Germany}
\author{Eva Monroy}
\affiliation{University Grenoble-Alpes, 38000 Grenoble, France}
\alsoaffiliation{CEA-Grenoble, INAC-PHELIQS, 17 av. des Martyrs, 38000 Grenoble, France}

\begin{document}

\begin{abstract}

We have characterized the photodetection capabilities of single GaN nanowires incorporating 20 periods of AlN/GaN:Ge axial heterostructures enveloped in an AlN shell. Transmission electron microscopy confirms the absence of an additional GaN shell around the heterostructures. In the absence of a surface conduction channel, the incorporation of the heterostructure leads to a decrease of the dark current and an increase of the photosensitivity. A significant dispersion in the magnitude of dark currents for different single nanowires is attributed to the coalescence of nanowires with displaced nanodisks, reducing the effective length of the heterostructure. A larger number of active nanodisks and AlN barriers in the current path results in lower dark current and higher photosensitivity, and improves the sensitivity of the nanowire to variations in the illumination intensity (improved linearity). Additionally, we observe a persistence of the photocurrent, which is attributed to a change of the resistance of the overall structure, particularly the GaN stem and cap sections. In consequence, the time response is rather independent of the dark current.

\end{abstract}

\vspace{10mm}

Semiconductor nanowires have received significant attention in the past few years as the miniaturization of electronic circuits and optoelectronic components continues.\cite{Li_2006,Lu_2007,Yan_2009,Zhang_2015a} Particularly, the use of nanowires as emitters and detectors in integrated photonic circuits for on-chip optical communication has been demonstrated recently.\cite{Tchernycheva_2014} Both single nanowires and nanowire arrays are studied as potential photodetectors using a variety of materials,\cite{VJ_2011} among them (Al,Ga)As,\cite{Dai_2014,Erhard_2015} ZnSe,\cite{Oksenberg_2015} ZnTe,\cite{Cao_2011} ZnO\cite{Cheng_2013,Zhang_2013} and (Al,Ga)N.\cite{Rigutti_2010, Gonzalez-Posada_2012, Gonzalez-Posada_2013, Jacopin_2014, Chen_2014} The high crystalline quality of nanowires, their low dimensionality and their high surface to volume ratio are important elements to understand the performance of nanowire photodetectors and identify their potential advantages. In general, nanowires present high photocurrent gain, in the range of $10^5-10^8$, with the photocurrent scaling sublinearly with the excitation power, and a time response of the photocurrent in the millisecond range.\cite{Gonzalez-Posada_2012} In comparison with planar devices, almost defect-free semiconductor nanowires open the possibility of exploiting the photoconductive gain while avoiding the deleterious effects of grain boundaries and dislocations on the spectral response.\cite{delaMata_2014} Both ZnO and (Al,Ga)N are promising choices to be used as miniature, visible-blind, ultraviolet (UV) photosensors,\cite{Cheng_2013,Zhang_2013,Rigutti_2010,denHertog_2012,Gonzalez-Posada_2012, Jacopin_2014} but GaN presents advantages in terms of physical and chemical robustness as well as for controlled doping.

The incorporation of axial heterostructures in wurtzite nanowires opens prospects for tuning the cut-off wavelength and improving the response at low bias as a result of the asymmetric potential profile generated by the polarization-induced internal electric field. Previous studies have shown that axial heterostructures reduce the dark current,\cite{Rigutti_2010} but the spontaneous, unintentional formation of a GaN shell around the heterostructure can generate a significant shunt conduction path.\cite{Rigutti_2010,denHertog_2012} This parallel conduction can be exploited for the fabrication of chemical sensors\cite{denHertog_2012}, but it is undesirable in GaN/AlN photodetectors since it reduces the environmental stability and masks the advantages of the inserted heterostructure. In the current study, we investigate AlN/GaN superlattices in nanowires without GaN shell. The incorporation of the heterostructure leads to a decrease of the dark current and an increase of the photosensitivity. A significant dispersion in the dark current is attributed to the coalescence of nanowires with displaced heterostructures. We present a systematic study of several nanowires with different dark current levels, which shows that the increase of the dark current leads to a decrease of the photosensitivity.

\begin{figure}[t!]
\centering
\includegraphics*[width=\columnwidth]{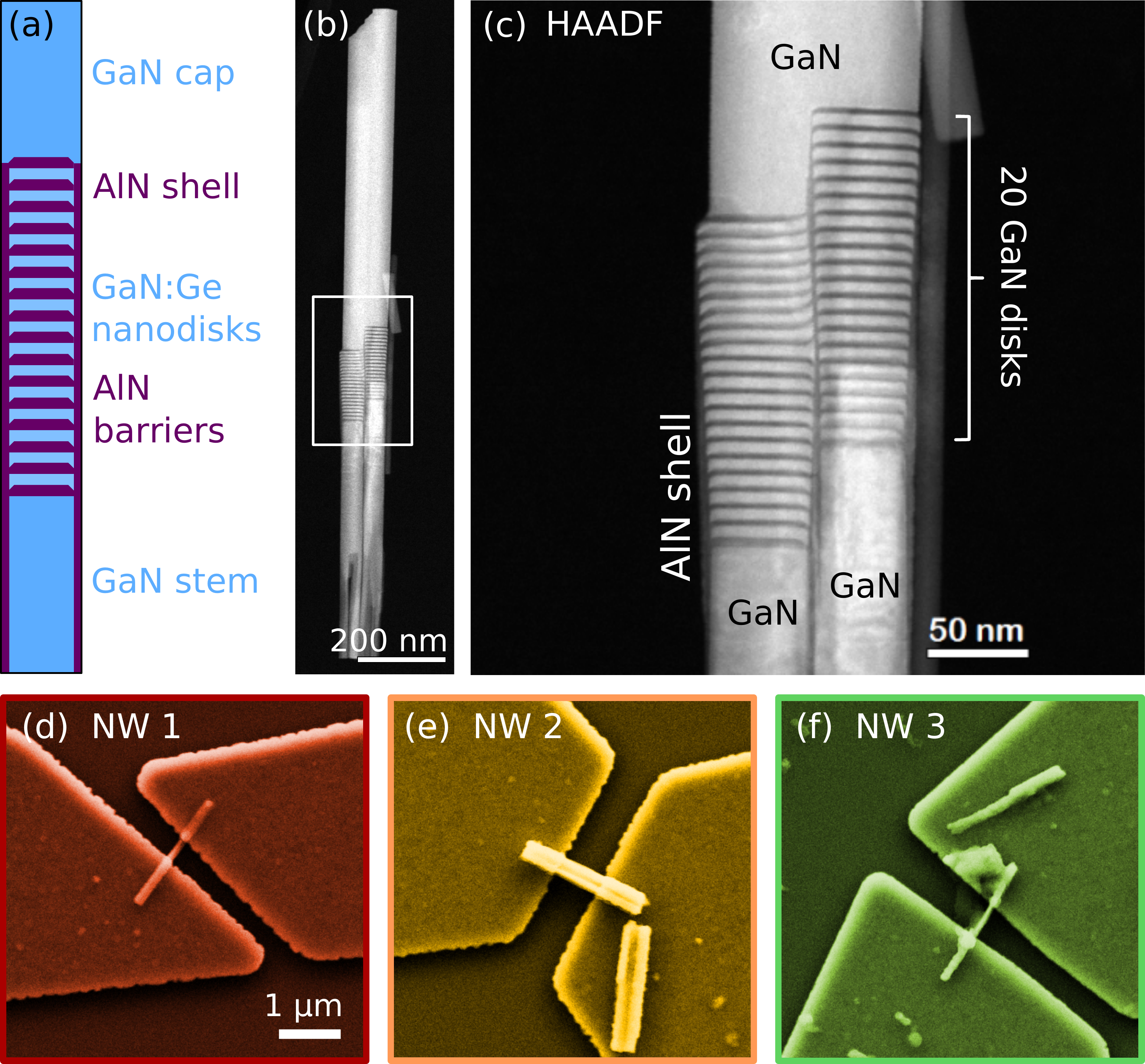}
\caption{\label{fig:tem}(a) Sketch of the nanowire structure under study. (b),(c) HAADF STEM images of two coalesced nanowires with different magnification. The bright and dark grey areas correspond to GaN and AlN, respectively. (d)--(f) SEM images of the contacted NWs~1--3, the same coloring as in the micrographs will be used in the following graphs to identify the individual nanostructures.}
\end{figure}


The nanowire heterostructures investigated in this study are schematically described in Figure~\ref{fig:tem}a. The growth of a GaN stem with a length of $\approx 600$~nm was followed by 20 periods of AlN barriers and GaN nanodisks. The nanodisks were doped with Ge at a concentration of $\approx2\times10^{20}$~cm$^{-3}$ (estimation from reference samples).\cite{Hille_2014} The heterostructure was capped by a $\approx 600$~nm long segment of GaN to facilitate contacting by electron beam lithography. The nanowire stem and cap are nominally undoped. The heterostructures were characterized by high-angle annular dark field (HAADF) scanning transmission electron microscopy (STEM) exemplified in Figures~\ref{fig:tem}b,c. The average thickness of the GaN nanodisks and AlN barriers were measured to be 5.3~nm and 2.7~nm, respectively. The gradient of the nanodisk thickness along the heterostructure is negligible. The chemical contrast in the HAADF micrographs of Figure~\ref{fig:tem}c reveals the presence of an AlN shell around the nanowire stem and the nanodisks, which results from lateral growth when depositing the AlN barriers.\cite{Furtmayr_2011a} At the same time, this micrograph and similar images of other nanowires (see Supporting Information) confirm the absence of a GaN shell that could form during the growth of the GaN cap. A significant number of nanowires coalesce during the growth, as commonly observed in self-assembled GaN nanowire ensembles.\cite{Brandt_2014} This coalescence can lead to the joining of nanowires with displaced nanodisk superlattices as seen in Figure \ref{fig:tem}c. The displacement is related to the length distribution that results from the statistical nature of the nanowire nucleation.\cite{Consonni_2013}

To investigate the photocurrent characteristics of individual nanowires, nanowires dispersed on SiN$_x$ were contacted using electron beam lithography. Three of the thus contacted nanowires, labeled NWs~1--3, are displayed in the scanning electron microscopy (SEM) images of Figures~\ref{fig:tem}d--f. The contacts are well centered, exposing the $\approx160$~nm long segment with the nanodisks. The nanowire diameters extracted from the SEM images indicate that NW~1 and NW~3 ($d\approx100$~nm) consist of at most two coalesced nanowires, whereas NW~2 consists of about 2--4 coalesced nanowires.


\begin{figure}[t!]
\centering
\includegraphics*[width=\columnwidth]{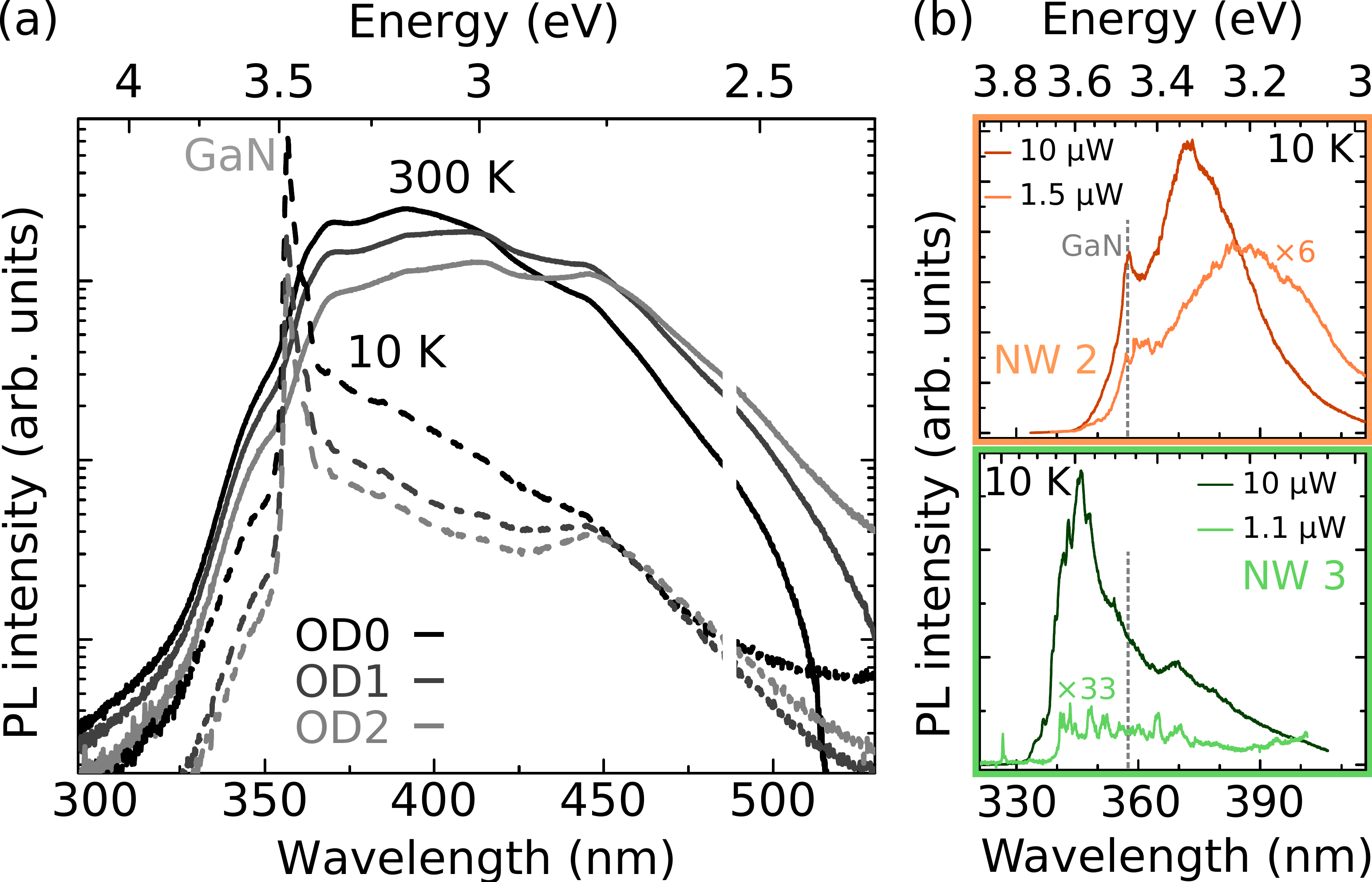}
\caption{\label{fig:pl}(a) PL spectra of densely dispersed nanowires measured at 10~K (dashed lines) and at 300~K (solid lines) at excitation powers of $200~\upmu$W (OD0), $12~\upmu$W (OD1) and $0.8~\upmu$W (OD2). At 488~nm, the second order of the laser peak has been masked. (b) Low-temperature micro-PL spectra of NWs~2 and 3 at two different excitation powers (linear intensity scale); the low excitation power spectra were multiplied by the indicated factors. Note that due to the difference in spot size, the excitation density corresponding to $10~\upmu$W in (b) is an order of magnitude higher than the strongest excitation density in (a). All spectra are normalized by the incident laser power.}
\end{figure}

To characterize the GaN/AlN heterostructures in the nanowires, photoluminescence (PL) spectra of nanowires mechanically
dispersed on a SiN$_x$-on-Si substrate, were collected under laser excitation at a wavelength $\lambda=244$~nm, with the laser focused to a spot of $\approx100~\upmu$m diameter. Figure~\ref{fig:pl}a shows the PL spectra from densely dispersed nanowires recorded for three excitation powers both at 10~K and 300~K. For normalization, the spectra are divided by the incident laser power. At low-temperature (10~K), the strongest PL feature is a narrow line at 3.472~eV assigned to the donor-bound exciton in the GaN stem and cap sections, with a low-energy shoulder at 3.42~eV related to emission of excitons bound to I$_1$ type stacking faults commonly observed in GaN nanowires.\cite{Lahnemann_2014} Additionally, a broad emission band extends roughly from 2.7 to 3.6~eV. At room temperature, the GaN band edge emission is strongly quenched,\cite{Hauswald_2014} and only the broad band remains, now extending from 2.5 to 3.6~eV. This band is assigned to emission from then GaN nanodisks embedded in AlN, since it is known that the three-dimensional confinement in the nanodisks hinders nonradiative recombination and hence leads to the persistence of the PL up to room temperature.\cite{Furtmayr_2011a,Beeler_2015} Independent of the temperature, an increase of the excitation power leads to an enhancement of the high-energy side of the nanodisk emission. This trend can be explained by both a saturation of the low-energy states in the nanodisks and a partial screening of the internal electric fields under high excitation densities.\cite{Lahnemann_2011}

The low-temperature PL spectra of two individual, contacted nanowires (NWs~2 and 3) are given in Figure~\ref{fig:pl}b, where the laser was focused to a spot of $\approx 3~\upmu$m diameter. For each nanowire, two excitation powers are shown. At high excitation power, the spectra consist of a $\approx 300$~meV broad emission band, which peaks at 3.33 and 3.58~eV for NW~2 and NW~3, respectively. The shift of almost 300~meV between the peak emission of both nanowires explains the broad emission of the nanowire ensemble, with a full width at half maximum $>500$~meV in Figure~\ref{fig:pl}a. With increasing excitation power, the spectra from single nanowires are blue-shifted as it was observed for the emission of the ensemble. For NW~3 (the nanowire emitting at higher energy), the PL peak decomposes into an ensemble of sharp lines at low excitation power. The sharp PL lines can be related to emission from different nanodisks\cite{Zagonel_2011} and to different excitation states within the nanodisks.

\begin{figure}[t!]
\centering
\includegraphics*[width=\columnwidth]{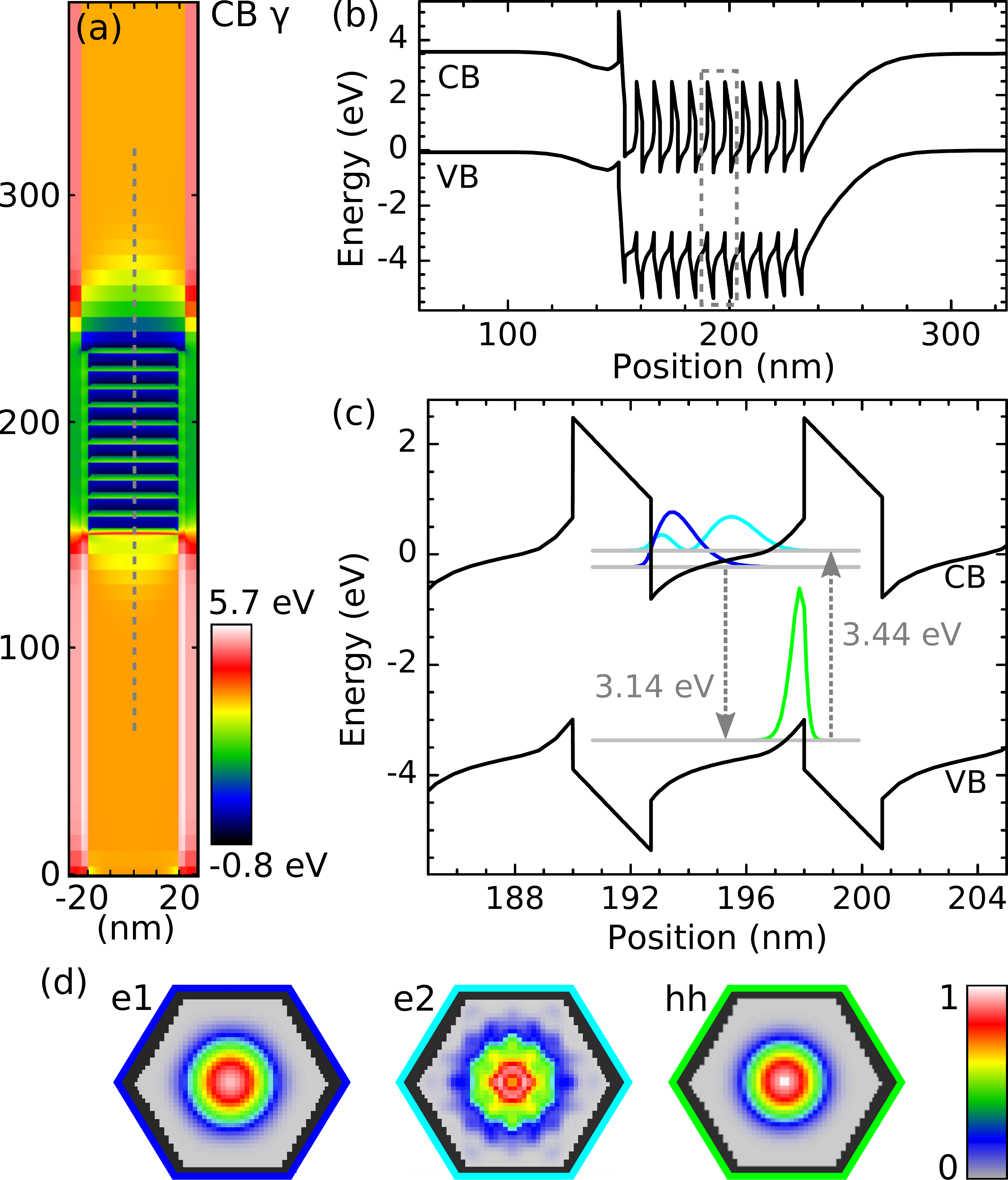}
\caption{\label{fig:nn}Results of three-dimensional Schr\"odinger-Poisson calculations. (a) Color-coded cross-sectional view of the conduction band edge. (b) Band profile along $\left[000\bar{1}\right]$ at the center of the nanowire [dashed line in (a)]. (c) Band profile along $\left[000\bar{1}\right]$ of a single nanodisk highlighting the transitions between the heavy-hole and the first and second electron levels, as well as the axial components of the squared wave functions [zoom of the marked area in (b)]. (d) In-plane cross-sectional view of the squared wave functions in (c) at their axial maximum (normalized color scale).}
\end{figure}

To understand the observed emission energy, we have performed three-dimensional calculations of the band diagram and confined levels in the nanodisk. A cross-sectional view of the calculated conduction band (CB) edge is shown in Figure~\ref{fig:nn}a. The conduction band and valence band (VB) profiles along $\left[000\bar{1}\right]$ in the center of the nanowire are given in Figure~\ref{fig:nn}b. The negatively-charged surface states lead to the depletion of the GaN stem and cap segments. In contrast, at the superlattice, the conduction band bends down and gets close to the Fermi-level (at 0~eV) due to the Ge-doping of the nanodisks. Figure~\ref{fig:nn}c illustrates the band profile along $\left[000\bar{1}\right]$ of a single nanodisk in the middle of the stack, including the squared wave functions and eigenenergies for the electron (e1) and heavy-hole (hh) ground states and the first excited electron state (e2) with a secondary node along the $\left[000\bar{1}\right]$ axis. Radially, all of these wave functions present a maximum in the center of the nanowire as depicted in Figure~\ref{fig:nn}d. In axial direction, the quantum-confined Stark effect is partially compensated by the high doping,\cite{Hille_2014} but still leads to a clear separation of the e1 and hh wave functions towards the bottom and top interface of the nanodisk, respectively. The resulting transition energy of this transition is 3.14~eV, which agrees well with the center of the PL band.

\begin{figure}[t]
\centering
\includegraphics*[width=7cm]{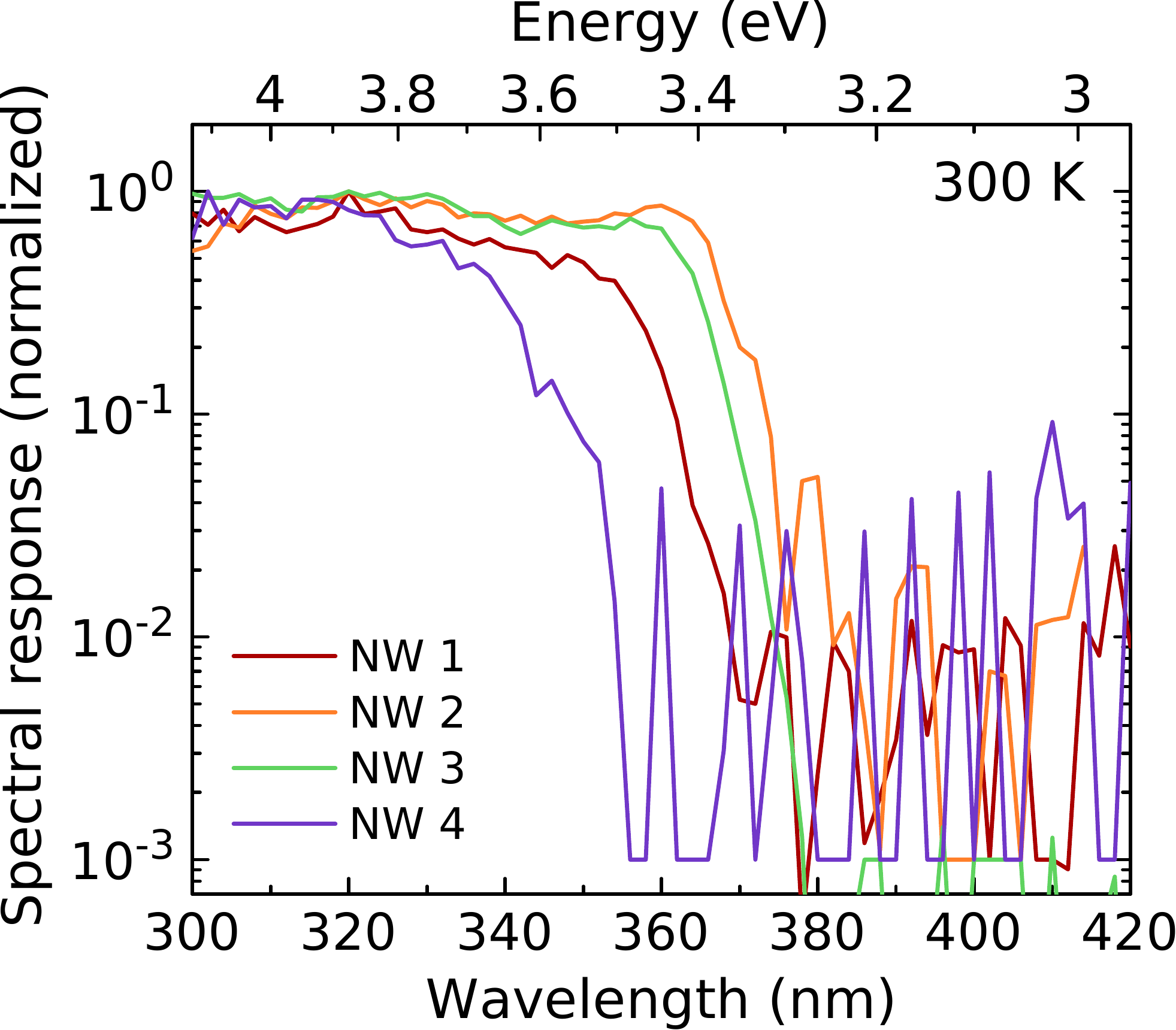}
\caption{\label{fig:sr}Normalized spectral response across the absorption edge for four different nanowires, among them NWs~1--3. The data are corrected by the spectral dependence of the lamp power and then normalized to their maximum.}
\end{figure}

In order to characterize the contacted single nanowires as photodetectors at room temperature, they were biased at 0.5--2~V, depending on the current-voltage ($I$--$V$) characteristics of the nanowires (Positive bias referring to the polarity giving the higher currents in the $I$--$V$ curves). The spectral responses of four different nanowires are presented in Figure~\ref{fig:sr}. They exhibit a sharp cut-off between 3.4 and 3.6~eV with the photocurrent dropping over 2--3 orders of magnitude, which corresponds to the detection limit of the setup. 
These spectral measurements unambiguously confirm that the photocurrent originates from the contacted nanowires, and not from the underlying silicon substrate. Furthermore, the variation in the energetic location of the spectral cut-off points towards the nanodisks playing the determining role in the photocurrent. The absorption edge agrees well with the hh--e2 transition at 3.44~eV determined from the simulations in Figure~\ref{fig:nn}. This blue shift of the photocurrent onset in comparison to the emission is due to the higher density of available states in the upper levels of the quantum well,\cite{Miller_1996} and the higher probability of escape of the photoexcited carriers. It is hence common that the transition observed by photoluminescence corresponds to e1--hh, whereas the dominant transition in the photocurrent spectra rather involves e2.\cite{Yu_1989,Parsons_1990} Moreover, in polar III-nitrides, the absorption involving excited states is further favored by the higher oscillator strength in comparison to hh--e1, which is a consequence of the quantum-confined Stark effect.

\begin{figure}[t]
\centering
\includegraphics*[width=7cm]{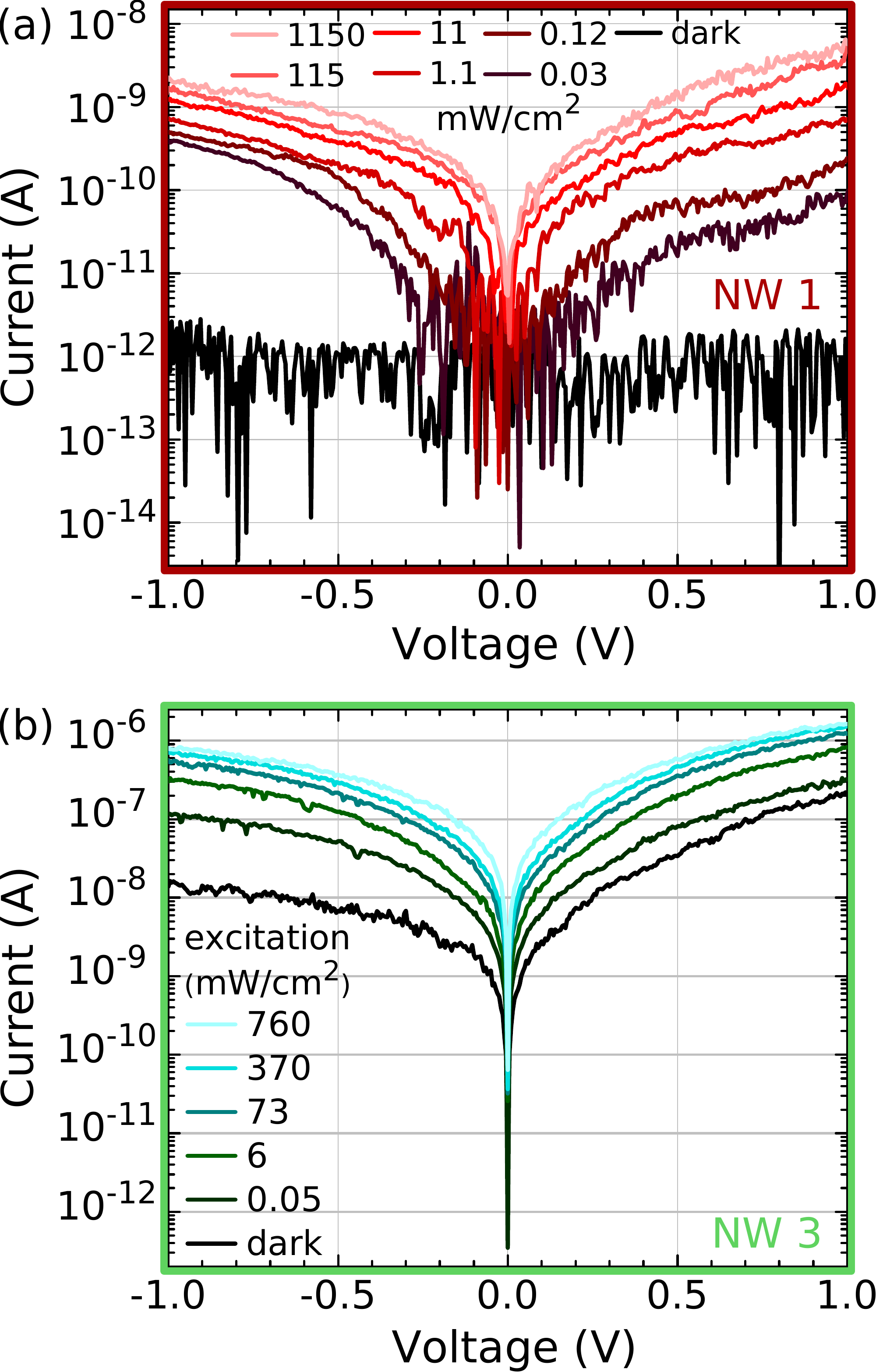}
\caption{\label{fig:i-v} $I$--$V$ characteristics of (a) NW~1 and (b) NW~3 measured in the dark and with increasing, continuous-wave UV laser ($\lambda=325$~nm) excitation density.}
\end{figure}

For a more detailed study of our nanowire photoresistors, the $I$--$V$ characteristics are investigated both in the dark and under continuous UV illumination (resulting in a direct current, DC) with a HeCd laser. There is a huge dispersion in the value of the dark current ($I_\mathrm{dark}$), which can vary from 2 pA to 0.6 $\upmu$A from wire to wire (values at 1~V bias). Figures~\ref{fig:i-v}a,b show the $I$--$V$ characteristics recorded on NWs~1 and 3, which are representative nanowires with low $I_\mathrm{dark}$ and high $I_\mathrm{dark}$, respectively. In both cases, they display a slightly asymmetric $I$--$V$ characteristic, consistent with the presence of the polarization-induced internal electric fields in the active region. With increasing UV illumination, the asymmetry is slightly reduced, as expected from the screening of the electric field at high carrier densities. 

\begin{figure}
\centering
\includegraphics*[width=7cm]{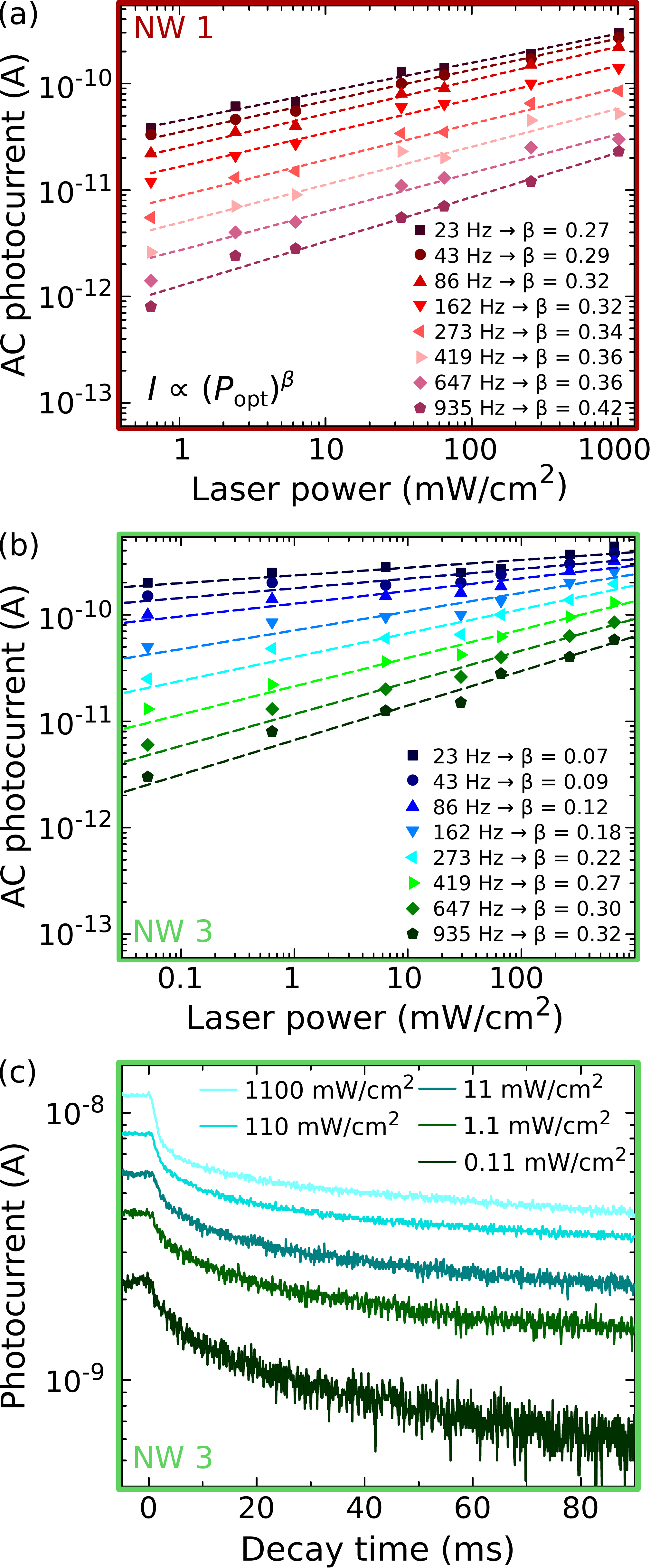}
\caption{\label{fig:freq}Photocurrent as a function of incident laser power $P_\mathrm{opt}$ at different chopping frequencies for (a) NW~1 (low $I_\mathrm{dark}$) and (b) NW~3 (high $I_\mathrm{dark}$) measured at 1 and 0.1~V bias, respectively. Dashed lines are fits by a sublinear power law with the exponents $\beta$ given in the legend. (c) Photocurrent decay of NW~3 for various excitation power densities measured at 0.5~V bias.}
\end{figure}

Additionally, the photogenerated signal has been studied as a function of the excitation power and the laser modulation frequency. Figures~\ref{fig:freq}a,b show the results for NW~1 (low $I_\mathrm{dark}$) and NW~3 (high $I_\mathrm{dark}$) measured at 1 and 0.1~V bias, respectively. The signal was recorded over a load resistance using lock-in detection (referred to as AC). Regardless of the chopping frequency, the nanowires show a sublinear response with the optical power of the laser $P_\mathrm{opt}$, which fits a power law with $I_\mathrm{ill} \propto (P_\mathrm{opt})^\beta$, where $I_\mathrm{ill}$ is the photocurrent, and $\beta < 1$ the sublinear exponent. For NW~3 , with high $I_\mathrm{dark}$, the sublinear behavior is very pronounced, with $\beta$ in the range of 0.07--0.32. For NW~1, having lower $I_\mathrm{dark}$, the values of $\beta$ are higher, in the range of 0.27--0.42, i.e.\ the latter nanowire is more sensitive to changes in the illumination intensity. 

The sublinear dependence of the photocurrent is commonly observed for nanowires, and not restricted to the III-nitride material system.\cite{Gonzalez-Posada_2012, Chen_2014, Cao_2011} In contrast to planar III-nitride heterostructures, where a similar behavior was attributed to extended defect states,\cite{Monroy_2003a} the sublinear scaling of the photocurrent in nanowires has been linked to surface states.\cite{Calarco_2005,Gonzalez-Posada_2012,Chen_2014} The neutralization of surface states by photogenerated holes leads to a reduction of the surface band bending and thus to the opening of a conductive channel in the center of an otherwise depleted nanowire. 

Additionally, the modulated photocurrent decreases markedly with increasing chopper frequency, which indicates a certain persistence of the photoconductivity. For low excitation densities (50~$\upmu$W/cm$^2$), the photocurrent decreases by up to two orders of magnitude when the chopping frequency increases from 23 to 935~Hz. This difference is reduced for excitation densities of $\approx1$~W/cm$^2$, which indicates an acceleration of the temporal response for high excitation densities. To confirm this point, we have also measured the photocurrent decay using an oscilloscope at a chopping frequency of 3~Hz. The results measured on NW~3 at 0.5~V bias for different excitation densities are depicted in Figure~\ref{fig:freq}c. The transients are strongly nonexponential. They consist of a fast initial decay that slows down after a few ms. The slope of the initial decay can be characterized by the $1/e$ time. In line with the measurements for varied chopping frequency, the decay time decreases from 13~ms to 5~ms when increasing the excitation density from 0.11~mW/cm$^2$ to 1.1~W/cm$^2$. Note that the measurements at 3~Hz (with a 50\% duty cycle) have a significant background, which amounts to about 20--40\% and indicates a persistence of the photoconductivity over longer time scales. In fact, in repeated $I$--$V$ measurements following exposure to the full intensity of the UV laser, it takes more than 20~minutes until the dark level of the $I$--$V$ characteristics is fully restored. An even stronger persistence of the photocurrent in III-nitride photoconductors is known from planar heterostructures, where it can take hours to restore the dark conductivity level.\cite{Monroy_2003a}

The photocurrent decays in Figure~\ref{fig:freq}c are orders of magnitude slower than the carrier lifetime in Ge-doped GaN nanodisks probed by time-resolved PL, which according to ref~\citenum{Beeler_2015} is around 100~ns for samples with a similar doping level. Therefore, the photogeneration of carriers in the nanodisks cannot lead to the persistence of the photocurrent. Instead, the photocurrent decay is comparable to that of simple n--i--n GaN nanowire photodetectors.\cite{Gonzalez-Posada_2012} In that case, similar to the sublinear response on incident illumination power, the persistent photoconductivity has been related to the charging of surface states combined with a radial separation of electrons and holes by the surface electric fields.\cite{Calarco_2005,Gonzalez-Posada_2012} However, the surface states should not affect the conductivity of the highly doped GaN nanodisks, where the band bending is restricted to a few nanometers close to the surface.\cite{Schuster_2015} The persistent phenomena are rather associated to the stem/cap GaN segments, which are fully depleted in the dark and where a conductive channel is opened under illumination. The decelerated discharging of surface states should thus delay the restoration of full depletion of these parts of the structure. Therefore, the persistence of the photocurrent is linked to a change in resistance of the overall structure, which modifies the current flowing under external bias. 

\begin{figure}[t!]
\centering
\includegraphics*[width=7cm]{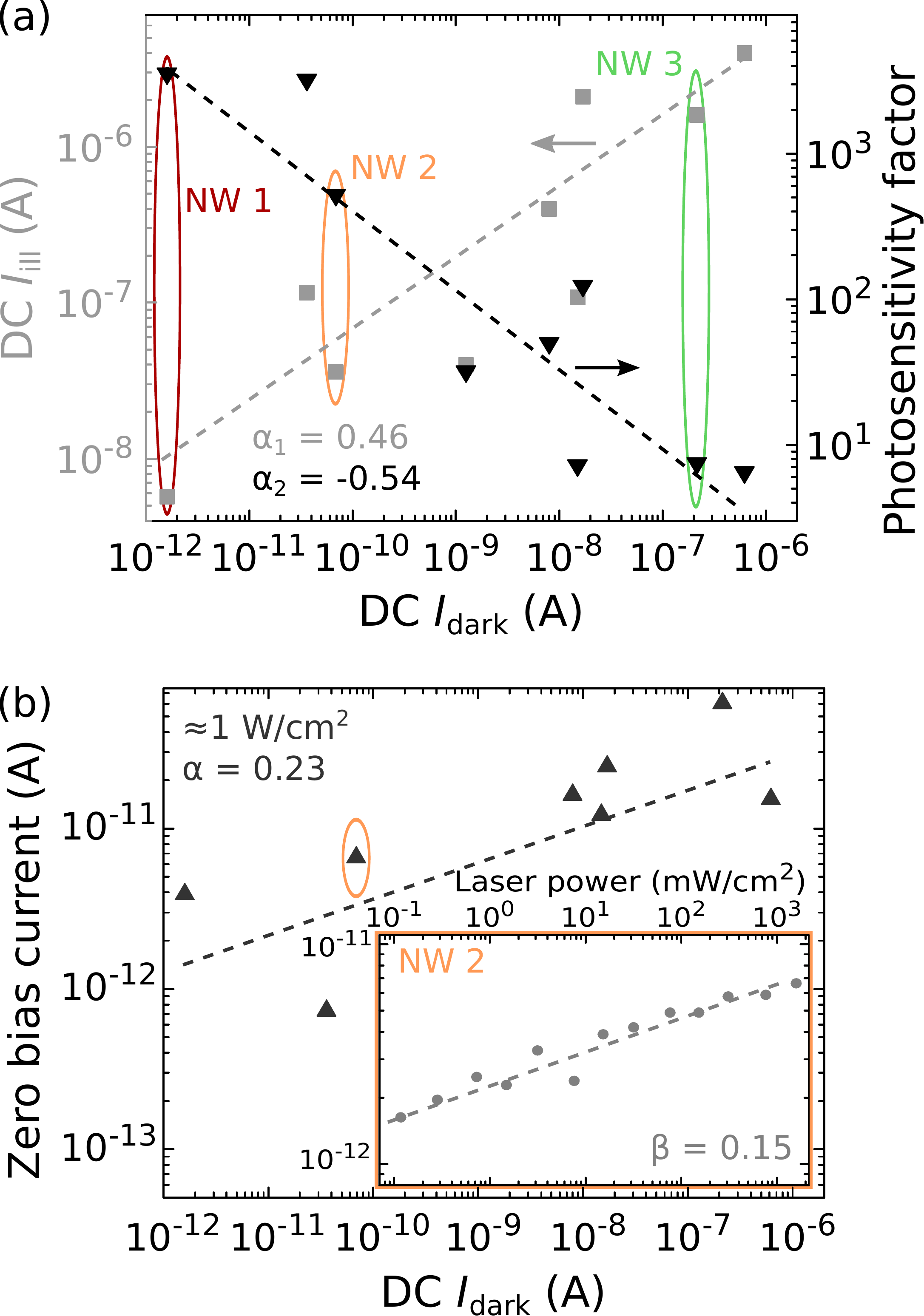}
\caption{\label{fig:stats} (a) $I_\mathrm{ill}$ and photosensitivity factor ($I_\mathrm{ill}/I_\mathrm{dark}$) for different nanowires plotted as a function of $I_\mathrm{dark}$. Both $I_\mathrm{ill}$ and $I_\mathrm{dark}$ are measured at 1~V bias. The value of $I_\mathrm{ill}$ corresponds to  an excitation density of $\approx 1$~W/cm$^2$ ($\lambda=325$~nm). (b) Photocurrent at zero bias for an excitation density of $\approx 1$~W/cm$^2$ ($\lambda=325$~nm) in different nanowires as a function of $I_\mathrm{dark}$ at 1~V bias. Dashed lines in (a) and (b) show fits with a sublinear power law and the respective exponents $\alpha$ (for dark current dependence) or $\beta$ (for illumination intensity dependence) are given. The inset in (b) shows the zero bias photocurrent of NW~2 as a function of the excitation density ($\lambda=325$~nm).}
\end{figure}

NWs~1 and 3 are representative examples of nanowires with high and low dark current. Looking for a correlation of $I_\mathrm{dark}$ with the nanowire properties, Figure~\ref{fig:stats}a presents a summary of characterization results obtained in nine single nanowires. For nanowires with increasing $I_\mathrm{dark}$, the photocurrent ($I_\mathrm{ill}$) at fixed bias and illumination level increases, though not as drastically as the dark current ($I_\mathrm{ill} \propto (I_\mathrm{dark})^{\alpha_1}$ with $\alpha_1=0.46$), which can lead to a decrease of the signal-to-noise ratio. The photosensitivity factor, defined by $I_\mathrm{ill}/I_\mathrm{dark}$ (both at 1~V bias), can be defined as a figure of merit. The value of  $I_\mathrm{ill}/I_\mathrm{dark}$, also plotted in Figure~\ref{fig:stats}a, decreases systematically as $I_\mathrm{dark}$ increases and follows a power law exponent $\alpha_2=\alpha_1-1=-0.54$. 

In previous studies, variations in $I_\mathrm{dark}$ between individual nanowires were attributed to the presence of a GaN shell of varying thickness around the heterostructures.\cite{Rigutti_2010,denHertog_2012} However, STEM measurements on the sample investigated in this study rule out the presence of such a shell, as shown in Figure~\ref{fig:tem} (see also Supporting Information). Instead, we observe that the coalescence of neighboring nanowires, associated with the formation of the AlN shell, can lead to several vertically displaced heterostructures within a single contacted nanowire (cf.\ Figure~\ref{fig:tem}c and Supporting Information). For displaced heterostructures, the current flows only through a part of the 20 superlattice periods (15 periods for the nanowire displayed in Figure~\ref{fig:tem}c), and thus the number of tunneling barriers changes between nanowires. The variation of this ``effective superlattice length'' can explain the observed differences in $I_\mathrm{dark}$ and confirms the reduction of $I_\mathrm{dark}$ by the introduction of the GaN/AlN heterostructure. In line with this discussion, the higher photosensitivity factors observed in nanowires with low $I_\mathrm{dark}$ are thus linked to a larger number of active superlattice periods in the structures. In contrast, the persistence of the photocurrent is mostly linked to a change of the resistance of the overall structure, in particular to the GaN stem and cap sections. Thus, the behavior is rather independent of $I_\mathrm{dark}$: the $1/e$ decay time for NW~3 at $\approx1$~W/cm$^2$ is 5~ms, to be compared to $\approx$ 18~ms for NW~2 (not shown), whose $I_\mathrm{dark}$ is more than three orders of magnitude smaller.

Alternatively to the displacement of the heterostructures, variations in the dark current level between single nanowires could be explained by differences in the nanowire diameters, but we found no correlation between the diameter measured on SEM images and the dark current.  Variations in the density of point defects could also have an impact on the dark current. However, if we assume a residual doping density\cite{Calarco_2005} of a few $10^{17}$~cm$^{-3}$ in the stem and cap segments of the nanowire, this corresponds to a few hundred donor atoms per segment. Assuming a Poisson distribution for the number of donors per segment, the average deviation from the mean for segments of the same volume is below 10\%, which does not justify conductivity changes by several orders of magnitude. Finally, a variable number of stacking faults at the base of the nanowires could also introduce a dispersion in the transport properties. However, the structural width and conduction band offsets associated to stacking faults, about 0.5--1.5~nm and 0.15--0.27~eV respectively,\cite{Lahnemann_prb_2012} are negligible in comparison to GaN/AlN heterostructures. In summary, though variations in the diameter, doping density or stacking fault density probably play a certain role, they are neither sufficient to explain the drastic (five orders) variations in dark current levels that we observe, nor could they account for the significant increase in photosensitivity at low dark current levels.

For a nanowire with misaligned contacts, where only the GaN stem (and not the superlattice) is contacted, the $I$--$V$ characteristics (not shown) are symmetric with $I_\mathrm{dark}=3~\upmu$A at 1~V bias, i.e., one order of magnitude higher than the highest value in Figure~\ref{fig:stats}a, whereas the photosensitivity is slightly lower than the smallest value in Figure~\ref{fig:stats}a. This observation is in line with measurements on n--i--n nanowire photodetectors in refs~\citenum{Rigutti_2010,Gonzalez-Posada_2012} and further confirms the beneficial effect of including the heterostructures.

Finally, all the nanowire heterostructures present a small photovoltaic response. The zero bias photocurrent (short-circuit current) is displayed in Figure~\ref{fig:stats}b, being of the order of several pA and scaling sublinearly with $I_\mathrm{dark}$ and excitation power (inset). This photovoltaic effect is linked to an asymmetry of the potential profile between the contacts, as previously observed in the case of asymmetric contacts on ZnO nanowires\cite{Liao_apl_2008} or in p-GaN/n-Si nanowire heterostructures.\cite{Tang_nl_2008} In our case, there are two main sources of asymmetry: (i) the polarization-related internal electric fields in the heterostructure, which could be further enhanced by reducing the doping level in the GaN nanodisks, and (ii) the asymmetry of stem and cap, the former being covered by a thin AlN shell. However, the photoresponse associated to a metal-insulator-semiconductor contact is expected to be linear with the excitation power.


In conclusion, the investigated Ge-doped AlN/GaN heterostructures in nanowires are suitable for nanoscale, visible-blind photodetector applications with threshold detection when linearity and speed are not essential. The spectral response clearly shows a sensitivity to wavelengths below about 360--380~nm, depending on the nanowire, and a rejection of longer wavelengths. This spectral behavior is consistent with three-dimensional calculations of the nanowire band-structure. As expected from the importance of surface states for the conductivity of nanowires, the photocurrent shows a sublinear dependence on the impinging optical power. This does not hinder the application for the detection of on--off situations. Persistent photoconductivity is observed, with an initial decay time on the order of a few-ms, but including also slower components.

Statistical measurements of dispersed nanowires from the same sample reveal a significant variation of their characteristics, with dark current levels varying over five orders of magnitude. This observation highlights the fact that conclusions should not be based on observations of only one or two nanowires from a given sample. Previous studies have associated the dark current to the spontaneous, non-intentional formation of a GaN shell around the heterostructure, which generated a non-negligible shunt conduction path. In the samples under study, the presence of a GaN shell has been discarded by HAADF-STEM studies. Instead, we relate the observed variation mainly to the coalescence of nanowires with displaced heterostructures. With different displacements, the ``effective superlattice length'' and thus the total barrier thickness in the nanowires changes. Lower dark currents correlate with higher photosensitivity factors, which is explained by a larger number of nanodisks contributing to the photocurrent. In contrast, the persistence of the photocurrent is mostly linked to a change of the resistance of the overall structure, and particularly to the GaN stem and cap sections. Thus, the time response is rather independent of the dark current.

The observed displacement is related to a length distribution of the base nanowires that results from the statistical nature of the nucleation process in self-induced nanowires.\cite{Consonni_2013} Therefore, to improve the homogeneity within a given sample, the length of the base nanowires prior to the growth of the heterostructures could be increased. With increasing growth time, the height distribution of self-assembled GaN nanowire arrays levels out and homogeneous lengths are achieved.\cite{Fernandez-Garrido_2015} Finally, note that the impact of displaced superlattices on the conductivity of coalesced nanowires is not specific to the investigated AlN/GaN photodetectors, but should be important also in nanowire based emitters and for other material systems.

\section{Experimental Methods}

The $\left[000\bar{1}\right]$-oriented nanowires investigated in this study were grown catalyst-free on Si(111) substrates by plasma-assisted molecular-beam epitaxy under N-rich conditions (Ga/N ratio $\approx 0.25$) and at a substrate temperature of $\approx 790$\,\celsius.\cite{Schormann_2013,Hille_2014} A periodic switching between Ga and Al fluxes leads to the formation of a superlattice. The high substrate temperature,\cite{Rigutti_2010} together with a shading effect from the high nanowire density and the widening during AlN growth, lead to a suppression of an additional GaN shell around the GaN/AlN superlattice. HAADF STEM was carried out using a FEI Tecnai F20 with a field emission gun operated at 200~kV, while SEM images were recorded in a Zeiss Ultra+ operated at 5~kV. For PL measurements, a continuous-wave solid-state laser ($\lambda = 244$~nm) was attenuated to the optical powers given in the text and used as excitation source. The light was focused to spots of $100~\upmu$m and $3~\upmu$m for ensemble and single nanowire measurements, respectively. The PL emission was dispersed in Jobin Yvon HR460 and Triax550 monochromators (for macro- and micro-PL, respectively) and detected with UV-enhanced charge-coupled device cameras.  In the as-grown nanowire ensemble, laser absorption in the thick GaN cap inhibits PL characterization of the GaN/AlN heterostructure. Thus, the ensemble PL measurements were performed on nanowires mechanically dispersed on a SiN$_x$-on-Si substrate.

To electronically contact individual nanowires, the as-grown nanowires are sonicated in solvent and the solution is then dispersed on n$^{++}$-Si substrates capped with $\approx 200$~nm of SiN$_x$. Contact pads and markers predefined by optical lithography facilitate the electrical contacting of single nanowires by electron beam lithography using a metal lift-off procedure. The deposition of 10~nm Ti and 120~nm Al was preceded by a 25~s argon plasma treatment to remove any residual photoresist and to improve the contact quality.

Three-dimensional calculations of the band structure in the nanowires were performed using the nextnano$^3$ software.\cite{Birner_2007} The parameters used in the calculations for GaN and AlN are summarized in a previous paper.\cite{Kandaswamy_2008} The nanowire was modeled as a hexahedral prism consisting of a 150~nm long GaN section followed by the AlN/GaN stack and capped with 150~nm of GaN. The nanodisks were defined similar as in ref~\citenum{Beeler_2015}, including the semipolar facets at the outer bottom interface of the nanodisks as sketched in Figure~\ref{fig:tem}a. The geometrical dimensions were taken from STEM measurements (core radius: 20~nm, AlN shell thickness: 3~nm, nanodisk thickness: 5.3~nm, barrier thickness: 2.7~nm). The n-type doping density in the nanodisks and the residual doping density were fixed to $2\times10^{20}$~cm$^{-3}$ and $5\times10^{17}$~cm$^{-3}$, respectively. The structure was defined on a GaN substrate to provide a reference in-plane lattice parameter, and was modeled as laterally embedded in a rectangular prism of air, which allowed elastic strain relaxation. In a first stage, the three-dimensional strain distribution was calculated by minimization of the elastic energy through the application of zero-stress boundary conditions at the surface. Then, for the calculation of the band profiles, the spontaneous polarization and the piezoelectric fields resulting from the strain distribution were taken into account. In lack of specific values for \emph{m}-plane AlN, the effect of surface states was simulated by introducing a two-dimensional negative charge density of $2\times10^{12}$~cm$^{-2}$ at the air/nanowire interface corresponding to the value reported for \emph{m}-plane GaN.\cite{Bertelli_2009} Wavefunctions and related eigenenergies of the electron and hole states in the nanodisks were calculated by solving the Schr\"odinger-Poisson equations using the effective mass approximation.

To measure the spectral response of nanowire photodetectors between 300 and 420~nm, they were excited with light from a 450~W Xenon lamp, passed through a grating monochromator, chopped at 86~Hz and focused to a spot of $\approx2$~mm diameter. The single nanowires were connected in series with a load resistance ($R=12$~M$\Omega$) and the photo-induced changes in the voltage across the load resistance were synchronously measured using a lock-in amplifier (Stanford Research Systems SR830). Measurements of the photocurrent as a function of illumination power and modulation (chopper) frequency were done using a similar setup with lock-in detection, but with illumination from a HeCd laser ($\lambda=325$~nm), the power of which was varied over four orders of magnitude using optical density filter wheels. A pinhole was used to reduce the spot diameter of the laser to $\approx 1$~mm. To record the photocurrent decay, the laser was modulated at 3~Hz and transients of the voltage drop across a load resistance were measured with an oscilloscope (Tektronix TDS 2022C). The $I$--$V$ characteristics were investigated both in the dark and under continuous illumination with the HeCd laser, again attenuated by optical density filter wheels to vary the excitation power (passing from low to high optical powers to avoid any alteration of the results by the persistent effects). The nanowires were directly connected to an Agilent 4155C semiconductor parameter analyzer and biased over a range of $\pm 1$~V. Positive bias in all these measurements is defined as the polarity giving the higher currents in the $I$--$V$ curves.

\section{Associated Content}
Supporting information: Additional transmission electron micrographs.

\begin{acknowledgement}
The authors would like to thank Bruno Gayral for help with the PL setup. Financial support from the EU ERC-SG ``TeraGaN'' (\#278428) and ANR JCJC COSMOS (ANR-12-JS10-0002) is acknowledged. Furthermore, the groups in Grenoble and Giessen received traveling support from the DAAD/CampusFrance program Procope. PH, JS and ME acknowledge financial support within the LOEWE program of excellence of the Federal State of Hessen (project initiative STORE-E). MdlM and JA acknowledge funding from Generalitat de Catalunya 2014 SGR 1638 and the Spanish MINECO MAT2014-51480-ERC (e-ATOM) and Severo Ochoa Excellence Program.
\end{acknowledgement}

\bibliography{UV-NW-PD.bib}

\end{document}